\documentclass[reprint,amsmath,amssymb,aps,prl,superscriptaddress,nofootinbib,floatfix]{revtex4-2}

\usepackage{amsmath,amssymb,amsthm,mathtools,bm}
\usepackage{graphicx}
\usepackage{float}
\usepackage{microtype}
\usepackage{dcolumn}
\usepackage{mathrsfs}
\usepackage{latexsym}
\usepackage{array}
\usepackage{multirow}
\usepackage{booktabs}
\usepackage{tikz}
\usetikzlibrary{patterns,decorations.pathmorphing,shapes}
\usepackage[colorlinks=true, linkcolor=blue, urlcolor=blue, citecolor=magenta]{hyperref}

\newtheorem{theorem}{Theorem}
\newtheorem{proposition}{Proposition}

\newtheorem{corollary}{Corollary}
\theoremstyle{definition}
\newtheorem{definition}{Definition}
\newtheorem{assumption}{Assumption}
\newtheorem{remark}{Remark}

\DeclareMathOperator{\Prb}{Pr}
\newcommand{\E}{\mathbb{E}}
\newcommand{\R}{\mathbb{R}}
\newcommand{\1}{\mathbf{1}}

\begin{document}

\title{From First Bond to Stable Adhesion: Multivalent Bond Stabilization in Receptor-Mediated Endocytosis}

\author{Hao Wu}
\email{wuhao@ucas.ac.cn}
\affiliation{Zhejiang Key Laboratory of Soft Matter Biomedical Materials, Wenzhou Institute,
University of Chinese Academy of Sciences, Wenzhou, Zhejiang 325000, China}

\date{\today}

\begin{abstract}
After a particle forms its first receptor--ligand bond near a cell membrane, stable attachment often requires a threshold total bond number \(m\). This paper develops a biophysical kinetic theory of this post-capture stage. The bond-number dynamics are modeled as a finite birth--death chain on \(n=0,1,\dots,m\), where \(0\) is loss of all bonds and \(m\) is the stabilization threshold. The mesoscopic rates \(\lambda_n\) and \(\mu_n\) encode the local energy landscape, accessible receptor--ligand pairs, steric and orientational constraints, and cooperative interactions within the growing cluster. The probability of reaching the threshold before losing all bonds is derived in closed form for general rates and, in the constant-rate case, reduces to a geometric series controlled by the kinetic bias ratio \(\rho=\mu/\lambda\). The mean time to failure or stabilization follows from a tridiagonal backward equation. Conditioning on successful attachment is treated as a trajectory-selection effect: the conditioned chain is a Doob \(h\)-transform whose dissociation rate at \(n=1\) vanishes, because successful trajectories cannot lose all bonds before stabilization. A minimal interface between the spatial first-capture probability and the bond-number chain is formulated, together with the conditions under which the stable capture probability factorizes into their product. The analysis shows that hitting probabilities constrain only the rate ratios, whereas the absolute rate scale requires time-resolved or dwell-time data. The threshold \(m\) is a model-defined stabilization criterion, not a directly measured ligand count. The rates and threshold are effective parameters specified independently or inferred from kinetic data. The theory provides a quantitative framework for interpreting attachment probabilities, lifetimes, and the transition from first bond formation to a model-defined stabilization threshold.
\end{abstract}

\maketitle

\section{Introduction}

Receptor-mediated endocytosis is a multi-stage process that underlies viral entry, nutrient uptake, and the internalization of engineered nanoparticles \cite{Doherty2009,zhang2015physical,wu2026unveiling,Gao2005,Shi2008,mercer2010virus,Marsh2006}. A particle first approaches the cell membrane through a crowded extracellular environment, then forms its first specific ligand--receptor bond, and finally the membrane wraps the particle while the cytoskeleton deforms and dissipates energy. Continuum models often treat the first approach as a boundary condition or assume that contact is already established. The pre-contact stage has been analyzed separately in a Smoluchowski--Robin framework in the companion paper \cite{Wu2026first}, where depletion attraction and near-wall diffusion determine the probability that a first bond forms before the particle escapes. The wrapping stage has also been analyzed separately in an Onsager variational framework, in which the ligand--receptor binding is compressed into a continuous adhesion energy density \(a\) \cite{Liu2026e1,Liu2026e2}. Between these two stages there is a discrete post-capture stage: after the first bond has formed, additional bonds form and break until the cluster either reaches a stabilization threshold or loses all bonds and fails. This stage is governed by bond-number fluctuations on a finite state space. It is not described by the pre-contact Smoluchowski--Robin theory, and it is not resolved by the continuous adhesion density of the wrapping theory. The present paper develops a kinetic theory of this intermediate stage.

The biological significance of this stage is direct. A single receptor--ligand bond is often too weak and too short-lived to commit a particle to internalization. Stable attachment typically requires a small cluster of bonds whose collective lifetime exceeds the time scale of membrane deformation and cytoskeletal reorganization. The transition from one bond to a stabilized cluster is therefore a decision point: if the cluster reaches the threshold before all bonds are lost, the particle proceeds toward wrapping; if all bonds are lost first, the attempt fails and the particle may escape or retry. This competition controls the fraction of successful attachment events and the distribution of attachment lifetimes, both of which are experimentally accessible. The same competition is central to viral infectivity, to the efficiency of targeted drug delivery, and to the design of multivalent nanoparticles. Related stochastic descriptions of adhesion clusters and bond-number fluctuations have been developed for shared-force and rebinding geometries \cite{Erdmann2004}; the present problem differs in its absorbing upper threshold, its first-capture interface, and its explicit conditioning on successful stabilization.

\begin{figure*}[t]
\centering
\includegraphics[width=\linewidth]{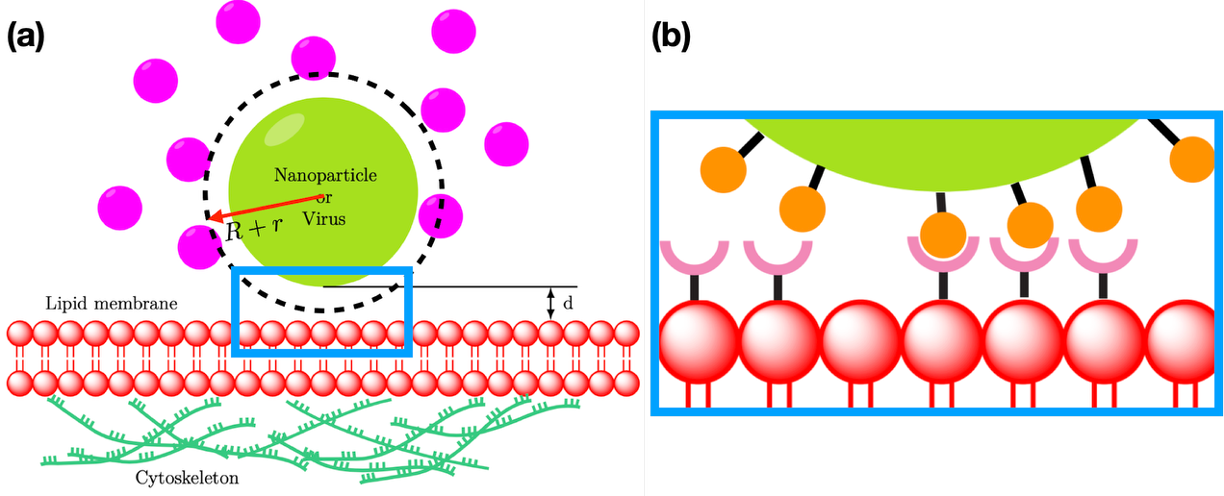}
\caption{Schematic of multivalent binding after first-bond formation. (a) A virus or nanoparticle approaches the cell membrane under depletion forces. (b) Zoom-in of the blue box in (a): multiple ligands on the virus or nanoparticle bind to multiple receptors on the cell membrane, illustrating the statistical process of bond formation after the first ligand--receptor complex has formed.}
\label{fig:schem}
\end{figure*}

The physical picture is as follows. Depletion forces near a wall drive the particle toward the membrane (see Fig.~\ref{fig:schem}(a)) and create a shallow potential well whose depth is of the order of a few \(k_{\mathrm B}T\) under typical crowding conditions \cite{Asakura1954,Asakura1958,MendezAlcaraz2000,Vrij1976,Dinsmore1995,Imhof1995,Steiner1995,Ilett1995,Dinsmore1998,Irajizad2017,Wan2024,Batra2009,Gupta2017,Dey2022,ellis2001crowding}. This well increases the residence time of the particle near the surface and therefore increases the probability that a first bond forms before the particle escapes \cite{Wu2026first}. Once the first bond has formed, the particle is no longer a simple point diffusing in an external potential (see Fig.~\ref{fig:schem}(b)). The bound receptor--ligand pair holds the particle at a preferred gap, and the remaining ligands and receptors fluctuate between associated and dissociated states. The bond number \(n\) becomes the relevant slow coordinate. The rates at which bonds form and break depend on the instantaneous separation, orientation, and the number of available partners. In a coarse-grained description, these microscopic details are absorbed into effective rates \(\lambda_n\) and \(\mu_n\). The resulting birth--death chain is the minimal Markovian model of the post-capture stage when the hidden configurational degrees of freedom relax faster than the bond-number coordinate and simultaneous multi-bond events can be neglected.

The association rate \(\lambda_n\) is proportional to the number of accessible unbound receptor--ligand pairs and to an intrinsic on-rate that may be reduced by steric and orientational constraints. The dissociation rate \(\mu_n\) is the thermal escape rate from the bound state, possibly modified by force or by cooperative interactions among bonds \cite{Bell1978,EvansRitchie1997}. The ratio \(\rho_n=\mu_n/\lambda_n\) is the kinetic bias ratio at fixed \(n\): it measures the relative probability of losing a bond versus gaining one at a given bond count. It is not, in general, the exponential of a single-bond free-energy difference. Under detailed balance for an underlying reversible chain, the free-energy difference between adjacent states involves \(\mu_{n+1}/\lambda_n\), not \(\mu_n/\lambda_n\). Only in the constant-rate model \(\lambda_n=\lambda\) and \(\mu_n=\mu\) can the common ratio \(\rho=\mu/\lambda\) be interpreted as a uniform kinetic bias per bond. The hitting probability \(p_1\), the probability that the chain reaches \(m\) before returning to \(0\), is the first physical observable of interest. The mean absorption time and the conditional stabilization time are the second and third observables. The distinction between probability and time is not a mathematical nicety: a chain can be very likely to stabilize eventually while taking a long time to do so, or it can stabilize quickly on the rare occasions when it succeeds. Experimental observables such as the fraction of stably attached particles and the distribution of attachment lifetimes probe different combinations of these quantities.

Multivalent binding is a cooperative thermodynamic effect in which the effective affinity of a cluster can exceed that of a single bond \cite{Kitov2003}. The cluster is not static: bonds form and break, and the distribution of bond numbers evolves in time. Recent work has emphasized the role of ligand mobility, configurational entropy, and steric interactions in the wrapping stage \cite{ZhangLiWang2021PRE,ChenXueLiEtAl2024,DiMicheleJanaMognetti2018}. These effects enter the present theory through the effective rates \(\lambda_n\) and \(\mu_n\), which are allowed to depend on \(n\). The birth--death chain does not resolve the microscopic origin of each rate, but it provides the exact probabilistic structure that any microscopic model satisfying the Markovian coarse-graining and time-scale separation must reduce to at the mesoscopic level.

The spatial first-capture problem and the bond-number chain are coupled in general. The local gap \(d\) affects the bond rates, and the number of bonds affects the effective potential and mobility. The fully coupled model is a Markov process on \((d,n)\) with a generator that contains both spatial diffusion and bond-number transitions \cite{Johnston2016,Billah2023,FedotovAlexandrov2024}. When the post-capture success probability is independent of the entry history and equals \(p_1\), the probability of stable capture factorizes into the spatial capture probability \(u(d_0)\) and the chain hitting probability \(p_1\). This paper states the conditions under which this factorization is physically justified and identifies the post-capture generator that must be embedded in a fully coupled model when it fails. The interface with the spatial stage is expressed through the capture probability \(u(d_0)\) of the Smoluchowski--Robin theory in the companion paper \cite{Wu2026first}, and the interface with the wrapping stage is expressed through the adhesion density \(a\) of the Onsager theory \cite{Liu2026e1,Liu2026e2}. The present paper therefore addresses the discrete layer between the two existing continuum descriptions.

The conditioning on successful stabilization is treated in the language of statistical mechanics. The event \(\{T_m<T_0\}\) selects a sub-ensemble of trajectories. The Doob \(h\)-transform gives the generator of this conditioned sub-ensemble \cite{Chetrite2015}. The transformed rates \(\lambda_n^*\) and \(\mu_n^*\) are not new microscopic rates; they describe the bias introduced by the selection. In particular, the conditioned chain has \(\mu_1^*=0\), meaning that under the success condition the total bond number cannot drop to zero before stabilization. This is a selection effect, not a modification of the underlying binding thermodynamics, and it does not mean that the particular bond formed first can never break. The conditional mean time \(T_n^*\) is the physically relevant time for experiments that observe only successful attachment events. The mathematical background for birth--death chains and first-passage problems is standard \cite{Norris1997,KarlinTaylor1975,Gardiner2004,Risken1989,Redner2001}, but its combination with a spatial capture interface and with a clear separation between probability and time has not been developed for the post-capture stage of multivalent adhesion.

The paper is organized as follows. Section~\hyperref[sec:model]{II} defines the birth--death chain and its generator. Section~\hyperref[sec:hitting]{III} derives the hitting probabilities and the constant-ratio closed form. Section~\hyperref[sec:mean-time]{IV} treats the mean absorption time. Section~\hyperref[sec:doob]{V} develops the Doob conditioning and the conditional stabilization time. Section~\hyperref[sec:spatial-bond]{VI} formulates the spatial--bond coupling interface and the factorization conditions. Section~\hyperref[sec:discussion]{VII} summarizes the physical interpretation, the model assumptions, and the identifiability conditions.

\section{The birth--death stabilization chain}
\label{sec:model}

\begin{definition}[Bond-number chain]\label{def:chain}
Let \(m\ge2\). The bond-number process \((N_t)_{t\ge0}\) is a continuous-time Markov chain on \(\{0,1,\dots,m\}\) with absorbing states \(0\) and \(m\). For \(1\le n\le m-1\),
\begin{equation}
\begin{aligned}
& n\longrightarrow n+1 \quad\text{at rate }\lambda_n>0,\\
& n\longrightarrow n-1 \quad\text{at rate }\mu_n\ge0.
\end{aligned}
\end{equation}
States \(0\) and \(m\) are absorbing. The process starts at \(N_0=i\) with \(1\le i\le m-1\) unless otherwise stated.
\end{definition}

The rate \(\lambda_n\) is the mesoscopic association rate at bond count \(n\). The rate \(\mu_n\) is the mesoscopic dissociation rate. Both rates have units of inverse time. They are not the microscopic on- and off-rates of a single receptor--ligand pair, because they include the multiplicity of available partners, steric constraints, and many-body effects. In a dilute ligand--receptor system with independent bonds, one would write \(\lambda_n\propto (N_{\mathrm{lig}}-n)(N_{\mathrm{rec}}-n)\) and \(\mu_n\propto n\), but such a factorization is a specific microscopic assumption and is not required by the general theory. The dependence of \(\lambda_n\) and \(\mu_n\) on \(n\) is the mesoscopic representation of multivalent cooperativity and of the changing local environment as the cluster grows \cite{Kitov2003,DiMicheleJanaMognetti2018}.

For a function \(f:\{0,\dots,m\}\to\R\), the generator acts on interior states as
\begin{equation}\label{eq:Q}
(\mathcal Q f)_n
=
\lambda_n\bigl(f_{n+1}-f_n\bigr)
+
\mu_n\bigl(f_{n-1}-f_n\bigr),
\ 1\le n\le m-1.
\end{equation}
Equivalently, on interior states the transition-rate matrix \(Q=(Q_{nn'})\) has
\[
Q_{nn}=-(\lambda_n+\mu_n),\
Q_{n,n+1}=\lambda_n,\
Q_{n,n-1}=\mu_n,
\]
with all other entries zero. The boundary values \(f_0\) and \(f_m\) are prescribed by the boundary problem at hand.

\begin{assumption}\label{ass:finite}
The chain is finite and \(\lambda_n>0\) for all \(1\le n\le m-1\). The dissociation rates satisfy \(\mu_n\ge0\). Under these conditions the interior states are transient and absorption at \(\{0,m\}\) occurs almost surely. If bidirectional reachability is required, one additionally assumes \(\mu_n>0\) for all \(1\le n\le m-1\); the general results below do not require this stronger condition.
\end{assumption}

The threshold \(m\) is a model-defined stabilization level. It represents the total number of bonds at which the cluster is considered stable for the purpose of the observation time window. Starting from \(n=1\), reaching \(m\) requires \(m-1\) net additional bonds. The threshold is not necessarily the total number of available ligands, and it may depend on the time resolution of the experiment. A cluster that is stable on the time scale of seconds may not be stable on the time scale of hours. The state \(0\) represents loss of all bonds and failure of the current attempt. A more detailed model could allow re-binding from \(0\), but that requires additional states or a renewal construction. The basic absorbing chain treats \(0\) as terminal for a single attempt.

\section{Hitting probabilities}\label{sec:hitting}

The first physical observable is the probability that the chain reaches \(m\) before returning to \(0\). Define
\begin{equation}\label{eq:hitting}
h_n:=\Prb_n(T_m<T_0),\quad 1\le n\le m-1,
\end{equation}
with boundary values
\[
h_0=0,\quad h_m=1.
\]
Here \(T_a=\inf\{t\ge0:N_t=a\}\).

\begin{theorem}[General hitting probability]\label{thm:hitting}
Under Assumption~\ref{ass:finite}, the hitting probabilities \(h_n\) are the unique solution of
\begin{equation}\label{eq:hitting-bvp}
(\mathcal Q h)_n=0,\ 1\le n\le m-1,
\ h_0=0,\ h_m=1.
\end{equation}
Equivalently,
\[
\lambda_n(h_{n+1}-h_n)+\mu_n(h_{n-1}-h_n)=0.
\]
Define the first differences
\[
\Delta_n:=h_n-h_{n-1},\qquad 1\le n\le m.
\]
Then
\begin{equation}\label{eq:Delta-recursion}
\Delta_{n+1}=\rho_n\Delta_n,
\qquad
\rho_n:=\frac{\mu_n}{\lambda_n}.
\end{equation}
Consequently,
\begin{equation}\label{eq:h_i-general}
h_i=
\frac{\displaystyle\sum_{k=0}^{i-1}\prod_{j=1}^{k}\rho_j}
{\displaystyle\sum_{k=0}^{m-1}\prod_{j=1}^{k}\rho_j},
\qquad 1\le i\le m-1,
\end{equation}
where the empty product is \(1\). In particular, starting from the first-bond state \(i=1\),
\begin{equation}\label{eq:p1-general}
p_1:=h_1=
\left[
1+\sum_{k=1}^{m-1}\prod_{j=1}^{k}\frac{\mu_j}{\lambda_j}
\right]^{-1}.
\end{equation}
\end{theorem}

\begin{proof}
The backward equation \((\mathcal Q h)_n=0\) follows from first-step analysis. In a short time \(\Delta t\), the chain stays at \(n\) with probability \(1-(\lambda_n+\mu_n)\Delta t+o(\Delta t)\), jumps to \(n+1\) with probability \(\lambda_n\Delta t+o(\Delta t)\), and jumps to \(n-1\) with probability \(\mu_n\Delta t+o(\Delta t)\). Thus
\[
h_n=
\bigl[1-(\lambda_n+\mu_n)\Delta t\bigr]h_n
+\lambda_n\Delta t\,h_{n+1}
+\mu_n\Delta t\,h_{n-1}
+o(\Delta t).
\]
Canceling \(h_n\) on both sides gives
\[
0=
-(\lambda_n+\mu_n)\Delta t\,h_n
+\lambda_n\Delta t\,h_{n+1}
+\mu_n\Delta t\,h_{n-1}
+o(\Delta t).
\]
Dividing by \(\Delta t\) and letting \(\Delta t\to0\) yields
\[
0=
-(\lambda_n+\mu_n)h_n
+\lambda_n h_{n+1}
+\mu_n h_{n-1}.
\]
Rearranging,
\[
\lambda_n(h_{n+1}-h_n)+\mu_n(h_{n-1}-h_n)=0,
\]
which is \eqref{eq:hitting-bvp}. For the differences, write \(h_{n+1}-h_n=\Delta_{n+1}\) and \(h_{n-1}-h_n=-\Delta_n\). Substitution gives
\[
\lambda_n\Delta_{n+1}-\mu_n\Delta_n=0,
\]
hence
\[
\Delta_{n+1}=\frac{\mu_n}{\lambda_n}\Delta_n=\rho_n\Delta_n,
\]
which is \eqref{eq:Delta-recursion}. Iterating this relation,
\[
\Delta_{k+1}=\left(\prod_{j=1}^{k}\rho_j\right)\Delta_1,
\qquad k=0,1,\dots,m-1,
\]
where the empty product for \(k=0\) equals \(1\). Since \(h_0=0\),
\[
h_i=\sum_{k=0}^{i-1}\Delta_{k+1}
=\Delta_1\sum_{k=0}^{i-1}\prod_{j=1}^{k}\rho_j.
\]
The condition \(h_m=1\) gives
\[
1=h_m=\Delta_1\sum_{k=0}^{m-1}\prod_{j=1}^{k}\rho_j,
\]
so
\[
\Delta_1=\left(\sum_{k=0}^{m-1}\prod_{j=1}^{k}\rho_j\right)^{-1}.
\]
Substituting this expression into the formula for \(h_i\) yields \eqref{eq:h_i-general}. Setting \(i=1\) gives
\[
p_1=h_1=\Delta_1=\left(\sum_{k=0}^{m-1}\prod_{j=1}^{k}\rho_j\right)^{-1},
\]
which is \eqref{eq:p1-general}, since the \(k=0\) term in the sum equals \(1\). Uniqueness follows from the maximum principle for the finite tridiagonal system.
\end{proof}

The ratio \(\rho_n=\mu_n/\lambda_n\) is the key kinetic parameter. It is the ratio of the dissociation rate to the association rate at bond count \(n\). It controls the hitting probability exactly, but it is not by itself a thermodynamic free-energy difference. Under detailed balance for an underlying reversible chain, the equilibrium weights \(\pi_n\) satisfy \(\pi_n\lambda_n=\pi_{n+1}\mu_{n+1}\), so the adjacent-state free-energy difference involves \(\mu_{n+1}/\lambda_n\). The product \(\prod_{j=1}^{k}\rho_j\) is a product of same-state kinetic biases, not a product of adjacent-state Boltzmann factors unless the rates are state-independent. The hitting probability \(p_1\) is a weighted sum over paths in this kinetic landscape.

\begin{figure}[t]
\centering
\includegraphics[width=\linewidth]{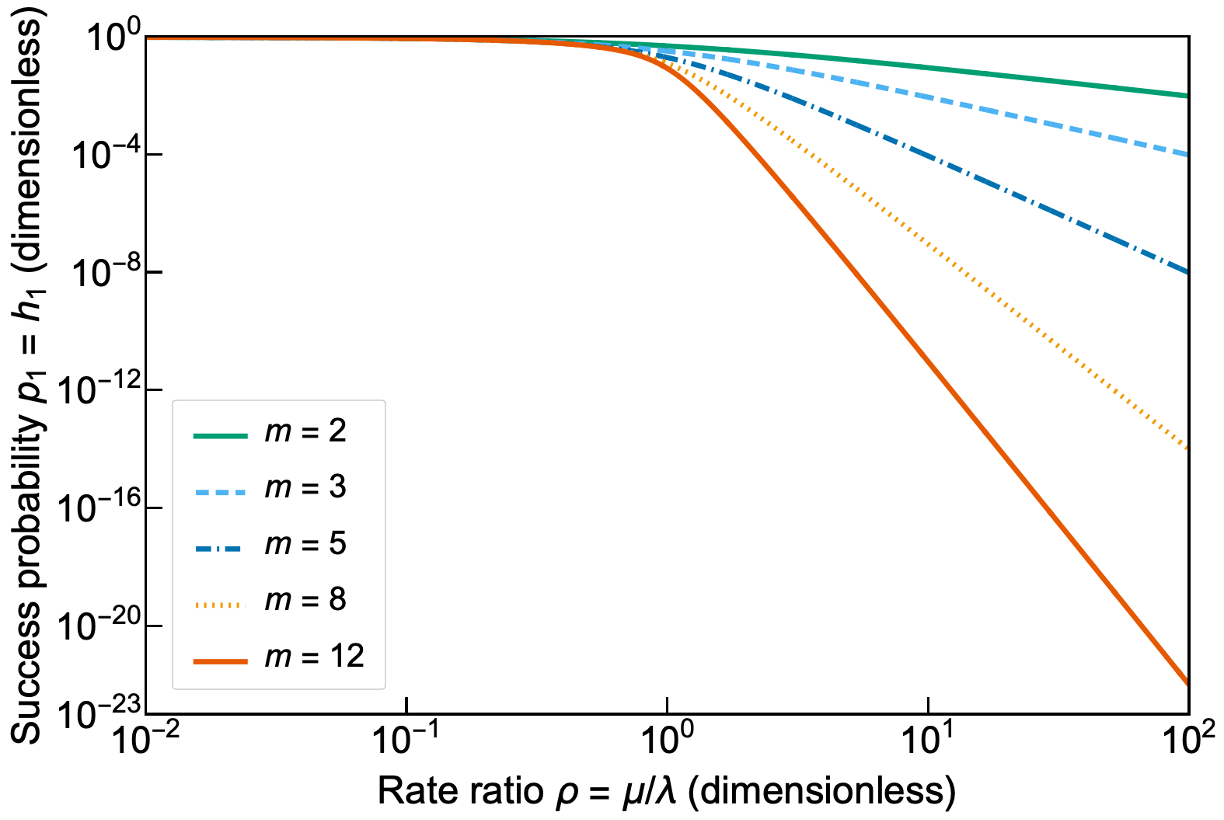}
\caption{Probability \(p_1\) that a constant-rate birth--death chain starting from the first-bond state reaches the threshold \(m\) before returning to \(0\). The rates are state-independent, \(\lambda_n=\lambda\) and \(\mu_n=\mu\), and the horizontal axis is the kinetic bias ratio \(\rho=\mu/\lambda\). The vertical axis is \(p_1=h_1\). Different curves correspond to different stabilization thresholds \(m\). The displayed range covers \(10^{-2}\le\rho\le10^2\).}
\label{fig:birth-death}
\end{figure}

\begin{corollary}[Constant-rate chain]\label{cor:constant-ratio}
Assume \(\lambda_n=\lambda>0\) and \(\mu_n=\mu\ge0\) for all \(1\le n\le m-1\), so that \(\rho_n=\rho=\mu/\lambda\). Then for \(\rho\ne1\),
\begin{equation}\label{eq:constant-ratio-pi}
h_i=\frac{1-\rho^i}{1-\rho^m},
\qquad 1\le i\le m-1,
\end{equation}
and in particular
\begin{equation}\label{eq:constant-ratio-p1}
p_1=\frac{1-\rho}{1-\rho^m}.
\end{equation}
For \(\rho=1\),
\[
h_i=\frac{i}{m},\qquad p_1=\frac{1}{m}.
\]
Moreover,
\[
\lim_{\rho\to0^+}p_1=1,
\qquad
\lim_{\rho\to\infty}p_1=0,
\]
and for fixed \(0<\rho<1\),
\[
\lim_{m\to\infty}p_1=1-\rho.
\]
\end{corollary}

\begin{proof}
If \(\rho_n=\rho\), then
\[
\prod_{j=1}^{k}\rho_j=\rho^k.
\]
Hence
\[
h_i=\frac{\sum_{k=0}^{i-1}\rho^k}{\sum_{k=0}^{m-1}\rho^k}.
\]
For \(\rho\ne1\), the geometric sum formula gives
\[
\sum_{k=0}^{i-1}\rho^k=\frac{1-\rho^i}{1-\rho},
\qquad
\sum_{k=0}^{m-1}\rho^k=\frac{1-\rho^m}{1-\rho}.
\]
Dividing the first expression by the second yields \eqref{eq:constant-ratio-pi}. Setting \(i=1\) gives \eqref{eq:constant-ratio-p1}. For \(\rho=1\), the recursion \(\Delta_{n+1}=\Delta_n\) gives \(\Delta_n=\Delta_1\) for all \(n\), and \(h_m=m\Delta_1=1\), so \(\Delta_1=1/m\) and \(h_i=i/m\). The limits follow by direct substitution.
\end{proof}

The constant-rate case is the discrete analogue of a constant force bias in a one-dimensional Brownian motion. The parameter \(\rho\) is a uniform kinetic bias, not a general free-energy parameter. For \(\rho<1\), the chain is biased toward the stabilized state; for \(\rho>1\), it is biased toward failure. The limit \(p_1\to1-\rho\) as \(m\to\infty\) for \(\rho<1\) shows that even an infinitely high threshold does not guarantee stabilization: there is always a finite probability \(\rho\) of returning to zero before reaching \(m\).

Figure~\ref{fig:birth-death} displays \(p_1\) as a function of \(\rho\) for several thresholds \(m\). The monotone decrease in \(\rho\) reflects the physical expectation that a larger dissociation-to-association ratio reduces the probability of reaching the stabilized state. The threshold dependence is strongest near \(\rho=1\), where the chain is nearly unbiased and \(p_1\approx1/m\).

\begin{remark}[Numerical stability]
For large \(m\) or rate ratios spanning many orders of magnitude, direct evaluation of products \(\prod_{j=1}^{k}\rho_j\) can overflow or underflow. A stable implementation uses logarithms. Set
\[
s_0=0,\qquad s_k=\sum_{j=1}^{k}\log\rho_j,
\qquad \rho_j=\frac{\mu_j}{\lambda_j}.
\]
If some \(\mu_j=0\), then \(\rho_j=0\) and the corresponding product vanishes; such zero rates should be handled separately before taking logarithms. For positive rates, factor out \(s_{\max}=\max_{0\le k\le m-1}s_k\) and evaluate
\[
h_i=
\frac{\sum_{k=0}^{i-1}e^{s_k-s_{\max}}}
{\sum_{k=0}^{m-1}e^{s_k-s_{\max}}},
\]
which avoids overflow. For the constant-rate formula, use the series expansion near \(\rho=1\) or the exact value \(i/m\) at \(\rho=1\).
\end{remark}

\section{Mean absorption time at \texorpdfstring{$\{0,m\}$}{0,m}}\label{sec:mean-time}

The probability \(p_1\) answers which boundary is reached first. It does not answer how long the process takes to reach either boundary. The mean absorption time is defined by
\[
T_n:=\E_n[\tau_{\{0,m\}}],
\
\tau_{\{0,m\}}=\inf\{t\ge0:N_t\in\{0,m\}\},
\]
with \(T_0=T_m=0\). This is the unconditional mean time to either failure or stabilization. It is a physically distinct observable from \(p_1\).

\begin{theorem}[Tridiagonal backward system]\label{thm:mean-time}
The vector \(\bm T=(T_1,\dots,T_{m-1})^{\mathsf T}\) is the unique solution of
\begin{equation}\label{eq:mean-system}
Q\bm T=-\bm 1,
\end{equation}
where \(Q\) is the interior transition-rate matrix with
\[
Q_{nn}=-(\lambda_n+\mu_n),\
Q_{n,n+1}=\lambda_n,\
Q_{n,n-1}=\mu_n.
\]
Equivalently,
\begin{equation}\label{eq:mean-recurrence}
\begin{aligned}
\lambda_n(T_{n+1}-T_n)+\mu_n(T_{n-1}-T_n)=-1,
\\ 1\le n\le m-1,
\end{aligned}
\end{equation}
with \(T_0=T_m=0\).
\end{theorem}

\begin{proof}
First-step analysis gives
\begin{equation}
\begin{aligned}
T_n=&\Delta t+
\bigl[1-(\lambda_n+\mu_n)\Delta t\bigr]T_n
+\lambda_n\Delta t\,T_{n+1}\\
&+\mu_n\Delta t\,T_{n-1}
+o(\Delta t).
\end{aligned}
\end{equation}
Canceling \(T_n\) gives
\begin{equation}
\begin{aligned}
0=&\Delta t
-(\lambda_n+\mu_n)\Delta t\,T_n
+\lambda_n\Delta t\,T_{n+1} \\
&+\mu_n\Delta t\,T_{n-1}
+o(\Delta t).
\end{aligned}
\end{equation}
Dividing by \(\Delta t\) and letting \(\Delta t\to0\),
\[
0=1-(\lambda_n+\mu_n)T_n+\lambda_nT_{n+1}+\mu_nT_{n-1}.
\]
Rearranging yields \eqref{eq:mean-recurrence}. In matrix form this is \eqref{eq:mean-system}. Since the finite chain is transient on interior states and absorption occurs almost surely, the interior submatrix \(Q\) is invertible with eigenvalues of negative real part, so the solution is unique.
\end{proof}

The physical content of \(T_n\) is the time scale for the bond cluster to resolve into either failure or stabilization. It is a combination of the intrinsic bond kinetics and the entropic barrier associated with reaching the threshold.

\begin{proposition}[Difference recurrence for the mean time]\label{prop:mean-difference}
Let
\[
s_n:=T_n-T_{n-1},\qquad 1\le n\le m.
\]
Then
\begin{equation}\label{eq:s-recursion}
s_{n+1}=\rho_n s_n-\frac{1}{\lambda_n},
\qquad 1\le n\le m-1,
\end{equation}
where \(\rho_n=\mu_n/\lambda_n\). The free constant \(s_1\) is determined by
\begin{equation}\label{eq:s-sum}
\sum_{n=1}^{m}s_n=T_m-T_0=0.
\end{equation}
\end{proposition}

\begin{proof}
Substitute \(T_{n+1}-T_n=s_{n+1}\) and \(T_{n-1}-T_n=-s_n\) into \eqref{eq:mean-recurrence}:
\[
\lambda_n s_{n+1}-\mu_n s_n=-1.
\]
Divide both sides by \(\lambda_n\):
\[
s_{n+1}-\frac{\mu_n}{\lambda_n}s_n=-\frac{1}{\lambda_n}.
\]
Since \(\rho_n=\mu_n/\lambda_n\), this is \eqref{eq:s-recursion}. The telescoping identity
\[
T_m-T_0=\sum_{n=1}^{m}(T_n-T_{n-1})=\sum_{n=1}^{m}s_n
\]
together with \(T_0=T_m=0\) gives \eqref{eq:s-sum}.
\end{proof}

The difference recurrence shows that the local time increment \(s_n\) is driven by the inverse association rate \(1/\lambda_n\), and it is propagated by the ratio \(\rho_n\). A large \(\lambda_n\) makes the local time increment small; a large \(\mu_n\) makes the propagation strong.

\section{Conditional stabilization and Doob conditioning}\label{sec:doob}

\begin{figure*}[t]
\centering
\includegraphics[width=\linewidth]{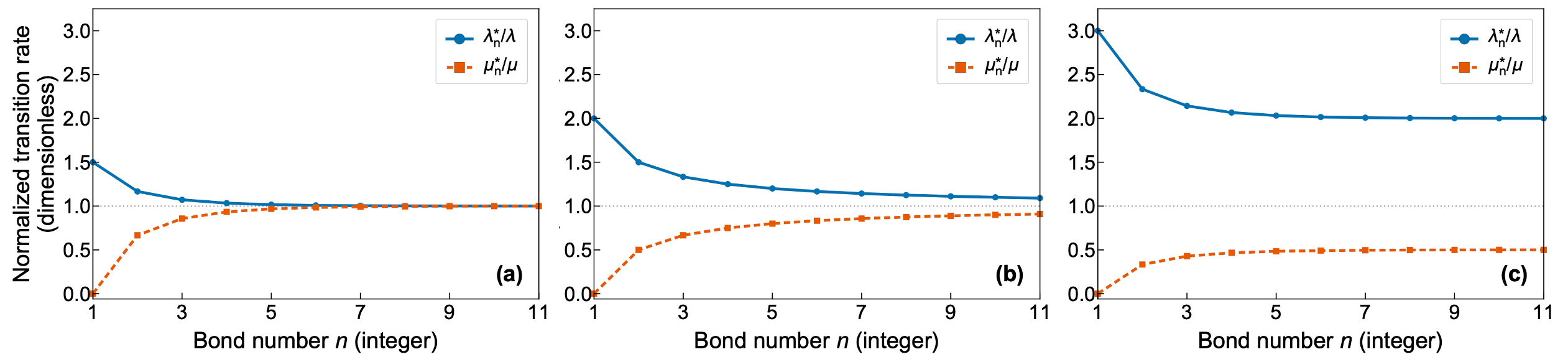}
\caption{Doob-transformed rates \(\lambda_n^*\) and \(\mu_n^*\) as functions of bond number \(n\) for a constant-rate chain with \(\lambda_n=\lambda\) and \(\mu_n=\mu=\rho\lambda\). The association rate \(\lambda_n^*\) is enhanced relative to \(\lambda_n\) because successful trajectories are biased toward higher bond numbers. The dissociation rate \(\mu_n^*\) vanishes at \(n=1\), reflecting the selection effect that the total bond number cannot reach zero under the success condition. This does not mean that a particular bond cannot break. The rates are normalized by the unconditioned \(\lambda\) and \(\mu\). Three parameter sets (a) \(\rho=0.5, m=12\), (b) \(\rho=1, m=12\), and (c) \(\rho=2, m=12\) are shown. The data are generated from the Doob transform of Theorem~\ref{thm:doob}. The total exit rate is preserved, \(\lambda_n^*+\mu_n^*=\lambda_n+\mu_n\).}
\label{fig:doob-rates}
\end{figure*}

Experiments often observe only the subset of trajectories that achieve stable attachment. Conditioning on \(\{T_m<T_0\}\) selects this subset. The Doob \(h\)-transform gives the generator of the conditioned chain \cite{Chetrite2015}.

\begin{theorem}[Doob transform of the birth--death chain]\label{thm:doob}
For \(1\le n\le m-1\), define
\[
\lambda_n^*:=\lambda_n\frac{h_{n+1}}{h_n},
\qquad
\mu_n^*:=\mu_n\frac{h_{n-1}}{h_n}.
\]
Then the chain conditioned on \(\{T_m<T_0\}\) is a birth--death chain on \(\{1,\dots,m\}\), absorbed at \(m\), with rates \(\lambda_n^*\) and \(\mu_n^*\). In particular,
\[
\mu_1^*=\mu_1\frac{h_0}{h_1}=0,
\]
so under the success condition the total bond number cannot drop to zero before stabilization. For any test function \(f\),
\begin{equation}
\begin{aligned}
(\mathcal Q^{(h)}f)_n
&:=
\frac{1}{h_n}\bigl(\mathcal Q(hf)\bigr)_n \\
&=
\lambda_n^*\bigl(f_{n+1}-f_n\bigr)
+
\mu_n^*\bigl(f_{n-1}-f_n\bigr).
\end{aligned}
\end{equation}
\end{theorem}

\begin{proof}
Since \(\mathcal Qh=0\), we have
\[
\lambda_n h_{n+1}+\mu_n h_{n-1}
=
(\lambda_n+\mu_n)h_n.
\]
Therefore
\begin{equation*}
\begin{aligned}
&\bigl(\mathcal Q(hf)\bigr)_n \\
=&
\lambda_n(h_{n+1}f_{n+1}-h_nf_n)
+\mu_n(h_{n-1}f_{n-1}-h_nf_n)\\
=&
h_n\left[
\lambda_n\frac{h_{n+1}}{h_n}(f_{n+1}-f_n)
+\mu_n\frac{h_{n-1}}{h_n}(f_{n-1}-f_n)
\right]\\
&\quad+
f_n\left[
\lambda_n h_{n+1}+\mu_n h_{n-1}
-(\lambda_n+\mu_n)h_n
\right].
\end{aligned}
\end{equation*}
The last bracket vanishes by \(\mathcal Qh=0\). Dividing the remaining expression by \(h_n\) gives
\begin{equation}
\begin{aligned}
\frac{1}{h_n}\bigl(\mathcal Q(hf)\bigr)_n
=&
\lambda_n\frac{h_{n+1}}{h_n}(f_{n+1}-f_n) \\
&+\mu_n\frac{h_{n-1}}{h_n}(f_{n-1}-f_n).
\end{aligned}
\end{equation}
This is the conditioned generator with rates \(\lambda_n^*\) and \(\mu_n^*\). At \(n=1\), \(h_0=0\), hence \(\mu_1^*=0\). The conditioned chain is absorbed at \(m\).
\end{proof}

The rate \(\mu_1^*=0\) is the mathematical expression of a physical selection effect. Under the success condition, trajectories that would have lost all bonds early are absent from the ensemble. The conditioned chain therefore appears to have a vanishing dissociation rate at \(n=1\). This is not a modification of the microscopic binding thermodynamics, and it does not mean that the first-formed bond cannot break; it means that the total bond number cannot reach zero. The conditioned chain also preserves the total exit rate at each interior state:
\[
\lambda_n^*+\mu_n^*=\lambda_n+\mu_n.
\]
Thus conditioning changes the jump directions and the occupation of paths, not the local waiting-time clock at a fixed state.

Figure~\ref{fig:doob-rates} displays the conditioned rates for a representative set of parameters. The vanishing of \(\mu_1^*\) is the clearest signature of the trajectory selection. The enhancement of \(\lambda_n^*\) at intermediate \(n\) reflects the fact that successful trajectories spend more time in states from which upward transitions are likely.

\begin{table*}[t]
\centering
\caption{The three bond-number quantities \(T_n\), \(h_n\), and \(T_n^*=g_n/h_n\), together with their interior equations and boundary conditions.}
\label{tab:three-quantities}
\begin{tabular*}{\textwidth}{@{\extracolsep{\fill}}cccc@{}}
\hline
Quantity & Interior equation & Boundary at \(0\) & Boundary at \(m\) \\
\hline
\(T_n\) & \(\mathcal Q T=-1\) & \(T_0=0\) & \(T_m=0\) \\
\(h_n\) & \(\mathcal Q h=0\) & \(h_0=0\) & \(h_m=1\) \\
\(T_n^*=g_n/h_n\) & \(\mathcal Q g=-h\) & \(g_0=0\) & \(g_m=0\) \\
\hline
\end{tabular*}
\end{table*}

\begin{proposition}[Conditional stabilization time]\label{prop:conditional-time}
Let
\[
g_n:=\E_n\bigl[T_m\1_{\{T_m<T_0\}}\bigr],
\qquad 1\le n\le m-1,
\]
with \(g_0=g_m=0\). Then
\begin{equation}\label{eq:g-bvp}
(\mathcal Q g)_n=-h_n,
\qquad 1\le n\le m-1,
\end{equation}
and the conditional mean stabilization time is
\begin{equation}\label{eq:conditional-time}
T_n^*:=\E_n[T_m\mid T_m<T_0]
=
\frac{g_n}{h_n},
\qquad 1\le n\le m-1.
\end{equation}
Equivalently, \(T_n^*\) solves the conditioned backward equation
\begin{equation}\label{eq:conditional-time-bvp}
(\mathcal Q^{(h)}T^*)_n=-1,
\qquad 1\le n\le m-1,
\end{equation}
with \(T_m^*=0\).
\end{proposition}

\begin{proof}
The defective moment \(g_n\) satisfies the backward equation \((\mathcal Q g)_n=-h_n\) with zero boundary values. This follows by first-step analysis applied to the pair consisting of the time and the success indicator, or by differentiating the Laplace transform of the joint law at zero. Since \(g_n=h_nT_n^*\), we have
\[
\bigl(\mathcal Q(hT^*)\bigr)_n=-h_n.
\]
Dividing by \(h_n\) and using Theorem~\ref{thm:doob} gives
\[
\frac{1}{h_n}\bigl(\mathcal Q(hT^*)\bigr)_n
=
(\mathcal Q^{(h)}T^*)_n
=
-1.
\]
This is \eqref{eq:conditional-time-bvp}. The boundary condition \(T_m^*=0\) follows from \(g_m=0\) and \(h_m=1\).
\end{proof}

The conditional time \(T_n^*\) is the physically relevant time scale for experiments that record only successful stabilization events. It is generally different from the unconditional mean time \(T_n\), because the conditioning removes the trajectories that fail early.

\begin{figure}[t]
\centering
\includegraphics[width=\linewidth]{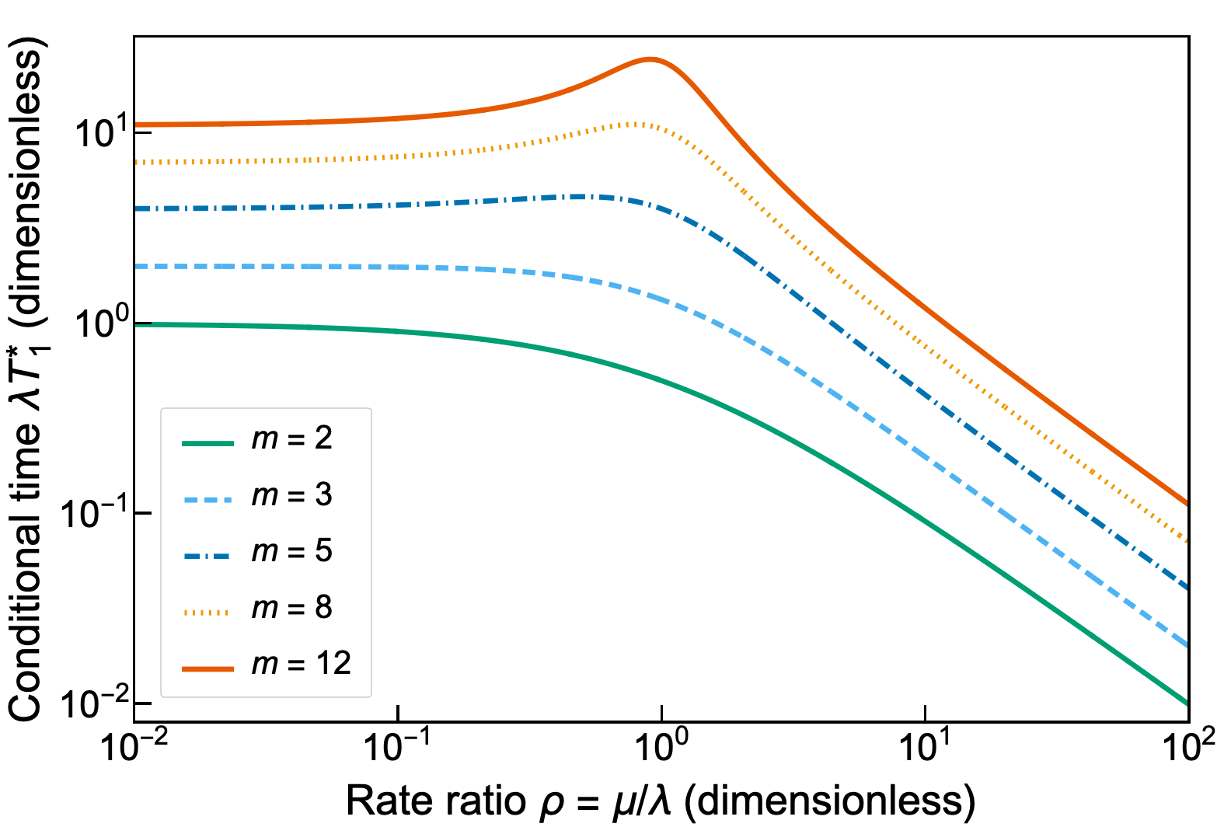}
\caption{Conditional mean stabilization time \(T_1^*=\E_1[T_m\mid T_m<T_0]\) as a function of the kinetic bias ratio \(\rho=\mu/\lambda\) for several thresholds \(m\), in the constant-rate model \(\lambda_n=\lambda\) and \(\mu_n=\rho\lambda\). The vertical axis is time in units of \(1/\lambda\). For small \(\rho\), the chain is strongly biased toward the stabilized state and \(\lambda T_1^*\to m-1\). At \(\rho=1\), \(\lambda T_1^*=(m^2-1)/6\). For large \(\rho\), conditioning on success forces the chain to take rapid upward steps against a strong dissociation bias, and \(\lambda T_1^*\sim(m-1)/\rho\). For \(m\ge4\) an interior maximum can occur near \(\rho<1\); for \(m=2,3\) the curve decreases monotonically with \(\rho\). The data are generated from Proposition~\ref{prop:conditional-time} with constant rates.}
\label{fig:conditional-time-rho}
\end{figure}

Figure~\ref{fig:conditional-time-rho} shows \(T_1^*\) as a function of \(\rho\). The small-\(\rho\) limit is linear in \(m\), \(\lambda T_1^*\to m-1\), while the unbiased point grows quadratically, \(\lambda T_1^*=(m^2-1)/6\). For large \(\rho\), the conditioned time decreases as \((m-1)/\rho\). For larger thresholds an interior maximum appears near \(\rho<1\); for \(m=2\) and \(m=3\) the curve decreases monotonically. Table~\ref{tab:three-quantities} records the equations and boundary conditions for the three quantities. The unconditional time \(T_n\) answers how long until either failure or stabilization; the hitting probability \(h_n\) answers which boundary is reached first; the conditional time \(T_n^*\) answers how long stabilization takes given that it occurs. The three quantities are related through \(g_n=h_nT_n^*\) but are not interchangeable. In particular, a chain can have high \(h_n\) and large \(T_n^*\), or low \(h_n\) and small \(T_n^*\).

\section{Spatial--bond interface and factorization conditions}\label{sec:spatial-bond}

The bond-number chain is coupled to the spatial coordinate \(d\), which measures the gap between the particle and the surface. Let \(d\in[a,b]\), where \(a=d_c\) is the capture distance and \(b=2r\) is the outer edge of the depletion layer. A fully coupled model must distinguish the pre-capture unbound state from the post-capture bound states. Let \(q_{\mathrm{pre}}(d)\) be the probability of stable capture starting from the unbound state at gap \(d\), and let \(q_n(d)\), \(1\le n\le m-1\), be the corresponding probability starting from \(n\) bonds at gap \(d\). The post-capture generator is
\begin{equation}\label{eq:post-generator}
\begin{aligned}
(\mathcal G f)_n(d)
=
&\mathcal L_n f_n(d)
+
\lambda_n(d)\bigl[f_{n+1}(d)-f_n(d)\bigr]\\
&+
\mu_n(d)\bigl[f_{n-1}(d)-f_n(d)\bigr],
\end{aligned}
\end{equation}
for \(1\le n\le m-1\), where \(\mathcal L_n\) is the spatial generator at fixed bond number \(n\). It includes the bond-number-dependent potential \(U_n(d)\) and mobility \(D_n(d)\),
\[
\mathcal L_n f
=
D_n f''+\bigl[D_n'-\beta D_n U_n'\bigr]f',
\]
together with the spatial boundary conditions appropriate to each \(n\). The rates \(\lambda_n(d)\) and \(\mu_n(d)\) depend on the local gap because bond formation and breakage depend on separation and orientation. Equation~\eqref{eq:post-generator} describes only the post-capture stage. It does not include first-bond formation, and it must not be interpreted as a complete coupled stabilization probability.

The spatial boundary conditions must distinguish the unbound pre-capture state from the bound states. For the unbound state, \(d=b\) is absorbing, because the particle can escape the depletion layer by non-specific diffusion. For every bound state \(n\ge1\), an intact bond tethers the particle to the membrane, so the particle cannot leave the near-wall region by the non-specific route that is available to an unbound particle. Two cases are then physically distinct. (i) If the tether formed by the \(n\) bonds has a contour length \(L_b\le b\), the particle cannot reach \(d=b\); the upper spatial boundary is effectively reflecting, and spatial escape at \(d=b\) is absent. (ii) If \(L_b>b\), the particle can reach \(d=b\) while still bound. The boundary at \(d=b\) is then partially absorbing, and the escape flux is controlled by the forced rupture of the stretched bonds rather than by free diffusion. A simple and physically transparent rule is
\begin{equation}\label{eq:escape-rule}
-D_n(b)\bigl[q_n'(b)+\beta U_n'(b)q_n(b)\bigr]
=
k_{\rm esc}(n)\,q_n(b),
\end{equation}
where \(k_{\rm esc}(n)\) is the force-dependent off-rate of the stretched cluster at \(d=b\) \cite{Bell1978,EvansRitchie1997}. Setting \(k_{\rm esc}(n)=0\) gives the reflecting limit of case (i), while \(k_{\rm esc}(n)\to\infty\) gives a perfectly absorbing boundary. At the inner boundary \(d=a\), \(q_n'(a)=0\) for \(n\ge1\), because a bound particle cannot cross the membrane; the reactive transfer below acts only on the unbound state.

A complete coupled model requires a pre-capture state and a transfer condition at the inner boundary. A simple and natural choice, in which first-bond formation moves the system into \(n=1\) at the same gap, is
\begin{equation}\label{eq:pre-transfer}
D(a)q_{\mathrm{pre}}'(a)
=
k_s\bigl[q_{\mathrm{pre}}(a)-q_1(a)\bigr],
\end{equation}
together with \(q_{\mathrm{pre}}(b)=0\) for the unbound state and the bound-state boundary rules specified above. Equation~\eqref{eq:pre-transfer} is one possible boundary transfer, not a unique consequence of the birth--death chain. If the first bond changes the gap or orientation, a transfer kernel replaces the point injection. If a fully dissociated cluster can rebind and restart, additional states or a renewal construction are needed. With these boundary rules, the coupled system is a well-posed starting point, but a complete quantitative model requires the rate functions \(\lambda_n(d)\), \(\mu_n(d)\), the escape rate \(k_{\rm esc}(n)\), and the transfer kernel to be fixed by microscopic input.

In the factorized approximation of Eq.~\eqref{eq:factorization}, the post-capture spatial escape is either negligible or already absorbed into the conditional success probability \(H(X_A)\). If neither holds, the coupled generator \eqref{eq:post-generator} with the boundary rule \eqref{eq:escape-rule} must be solved instead of using the product form.

\begin{figure}[t]
\centering
\includegraphics[width=\linewidth]{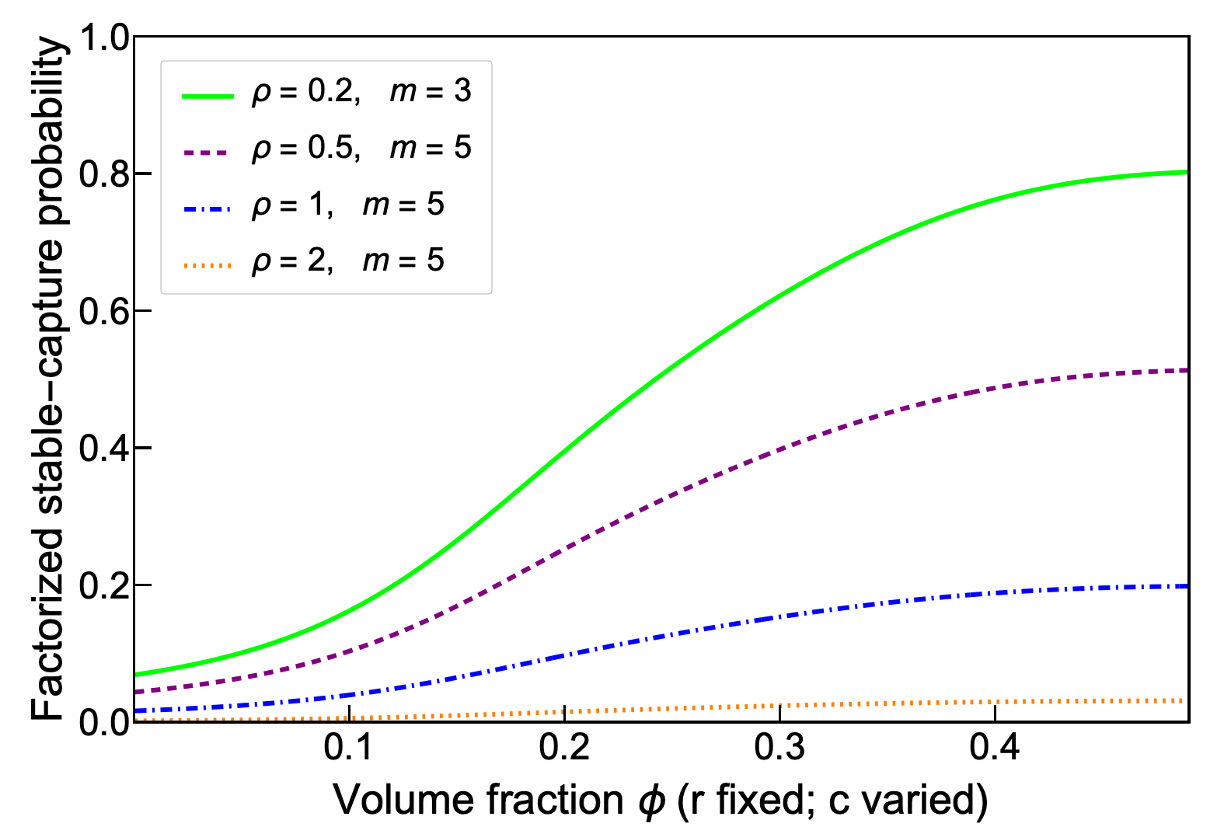}
\caption{First-bond reaction probability multiplied by the constant-rate birth--death hitting probability, at fixed crowder radius \(r\) and varying crowder concentration \(c\). The horizontal axis is the volume fraction \(\phi\); both axes are dimensionless. Different curves correspond to different conditional parameter pairs \((\rho,m)\). The spatial parameters are \(R=50\,\mathrm{nm}\), \(r=15\,\mathrm{nm}\), \(T=300\,\mathrm{K}\), \(\mu=10^{-3}\,\mathrm{Pa\,s}\), \(d_c=1\,\mathrm{nm}\), \(d_0=22.75\,\mathrm{nm}\), \(k_s=10^{-4}\,\mathrm{m\,s^{-1}}\), the CS pressure-only approximation, and \(P_{\mathrm{enter}}=1\); the near-wall mobility is \(D(d)=D_0d/(R+d)\). The curves cover the full displayed range up to the right frame edge at \(\phi=0.49\). The figure is meaningful only under the two-stage factorization assumptions of Sec.~\hyperref[sec:spatial-bond]{VI}.}
\label{fig:total-stabilization}
\end{figure}

When the post-capture success probability is independent of the entry history, the stable capture probability factorizes. Let \(A=\{\text{first bond forms before escape}\}\) and let \(X_A\) denote the configuration at first capture. Let \(H(X_A)\) be the probability of reaching \(m\) before losing all bonds, starting from that configuration. Then
\begin{equation}\label{eq:general-factorization}
P_{\mathrm{stable}}
=
P_{\mathrm{enter}}\,
u(d_0)\,
\E\bigl[H(X_A)\mid A\bigr],
\end{equation}
where \(u(d_0)\) is the spatial first-capture probability and \(P_{\mathrm{enter}}\) is the probability of entering the near-wall layer in the chosen protocol. The product form
\begin{equation}\label{eq:factorization}
P_{\mathrm{stable}}(d_0)
=
P_{\mathrm{enter}}\,u(d_0)\,p_1
\end{equation}
holds when \(H(X_A)=p_1\) almost surely under the conditional law given \(A\), or when its conditional mean has been independently shown to equal \(p_1\). The condition that the first bond does not appreciably change the potential or mobility is sufficient in many parameter regimes, but it is not the defining condition; the defining condition is the independence of the post-capture success probability from the entry history. If the initial gap has a distribution \(\rho_0(d)\), the spatial factor must be averaged,
\[
\int_a^b u(d)\rho_0(d)\,dd,
\]
and the same independence condition must hold for each entry configuration.

Figure~\ref{fig:total-stabilization} displays the product \(u(d_0)p_1\) under the factorization assumptions. The increase with \(\phi\) is driven mainly by the spatial capture probability \(u(d_0)\), which grows as the depletion well deepens. The chain factor \(p_1\) is independent of \(\phi\) in this plot because the rates are held fixed. When the rate-independence condition fails, the post-capture generator \eqref{eq:post-generator} must be embedded in a fully coupled model with the pre-capture transfer \eqref{eq:pre-transfer} or an appropriate generalization.

\begin{table}[t]
\centering
\caption{Mean reaction-before-escape probability \(\bar u\) and mean reflecting-outer MFPT \(\bar\tau\) for four conditional initial distributions of the spatial stage. All densities are normalized on \([a,b]\). The uniform density is \(1/(b-a)\). The restricted equilibrium density is the Boltzmann density of the reflecting reference system, \(\rho_{\mathrm{eq}}(d)\propto e^{-\beta U(d)}\), not the steady state of the reacting system. The entry-layer Gaussian has center \(b-\epsilon_{\mathrm{entry}}\) with \(\epsilon_{\mathrm{entry}}=0.5\,\mathrm{nm}\) and width \(\sigma=0.07L\), where \(L=b-a\). The near-capture Gaussian has center \(a+0.12L\) and the same width. Both Gaussians are truncated to \([a,b]\) and renormalized. These values are not parameters of the bond-number chain; they illustrate the dependence of the spatial factor \(u(d_0)\) on the initial preparation.}
\label{tab:initial-distributions}
\begin{tabular*}{\columnwidth}{@{\extracolsep{\fill}}ccc@{}}
\hline
Distribution & \(\bar u\) & \(\bar\tau\) (s) \\
\hline
Uniform & \(0.72023\) & \(1.57165\times10^{-4}\) \\
Restricted equilibrium & \(0.96739\) & \(8.28442\times10^{-5}\) \\
Entry-layer Gaussian & \(0.14000\) & \(2.16417\times10^{-4}\) \\
Near-capture Gaussian & \(0.97942\) & \(8.94867\times10^{-5}\) \\
\hline
\end{tabular*}
\end{table}

Table~\ref{tab:initial-distributions} shows that the spatial capture probability \(u(d_0)\) can vary by a factor of several depending on the initial distribution. The entry-layer Gaussian places most of the initial probability near the absorbing escape boundary and therefore gives the lowest \(u\) and the largest reflecting MFPT. The restricted equilibrium and near-capture distributions place more mass near the inner capture region and give the opposite trend. Because the product \eqref{eq:factorization} is linear in \(u(d_0)\), this uncertainty propagates directly to the stable capture probability. For this reason, any quantitative comparison of \(P_{\mathrm{stable}}\) with experiment must specify the initial distribution, just as it must specify \(k_s\), \(d_c\), and the bond-number rates.

\begin{proposition}[Time additivity under a fixed-restart structure]\label{prop:time-additivity}
Let \(T_A\) be the first-bond formation time and \(T_B\) the subsequent chain stabilization time. Conditional expectation is always linear:
\[
\E[T_A+T_B\mid A\cap B]
=
\E[T_A\mid A\cap B]
+
\E[T_B\mid A\cap B].
\]
To replace the joint conditioning by the separate events \(A\) and \(B\), one needs the selection-unbiased conditions
\[
\E[T_A\mid A\cap B]=\E[T_A\mid A],
\qquad
\E[T_B\mid A\cap B]=\E[T_B\mid B].
\]
A sufficient structural condition is a two-stage fixed-restart model: after \(A\), the post-capture chain starts from a fixed state \(n=1\) with a law independent of the full spatial history. Then \(T_B\) is independent of \(T_A\) given \(A\), and
\[
\E[T_A+T_B\mid A\cap B]
=
\E[T_A\mid A]+\E[T_B\mid B].
\]
Conditional independence of \(T_A\) and \(T_B\) given \(A\cap B\) alone is not sufficient to justify replacing the conditioning events. Without the selection-unbiased conditions or the fixed-restart structure, the right-hand side must retain the joint conditioning.
\end{proposition}

\begin{proof}
The first identity is the linearity of conditional expectation. The second identity follows from the selection-unbiased conditions. Under the fixed-restart model, \(T_B\) is generated by a chain with initial state \(n=1\) and law independent of the spatial history, so \(\E[T_B\mid A\cap B]=\E[T_B\mid B]\) and \(\E[T_A\mid A\cap B]=\E[T_A\mid A]\). The joint conditional expectation then separates. The final statement follows from the counterexample in which \(T_A\) and \(T_B\) are conditionally independent given \(A\cap B\) but the conditional means under \(A\cap B\) differ from those under \(A\) and \(B\) separately.
\end{proof}

\section{Discussion}\label{sec:discussion}

The birth--death chain provides a transparent physical model for the post-capture stage of multivalent adhesion. Its hitting probabilities have a closed product-sum form, and the constant-rate case reduces to a geometric series that is the discrete analogue of a constant-force Brownian barrier problem. The mean absorption time is obtained from a tridiagonal backward solve. Conditioning on successful stabilization yields a Doob-transformed chain in which the total bond number cannot reach zero before stabilization; the conditional mean time is the ratio of a defective first moment to the hitting probability. These results are standard in the mathematical theory of Markov chains \cite{Norris1997,KarlinTaylor1975}, and their combination with a spatial first-capture interface provides a reproducible kinetic description of multivalent adhesion. The contribution here is not the birth--death chain itself but the explicit interface with the spatial capture probability, the trajectory-selection analysis of conditioning on success, and the identifiability conditions for the rate parameters.

The physical interpretation of the main quantities is as follows. The ratio \(\rho_n=\mu_n/\lambda_n\) is the local kinetic bias per bond. The hitting probability \(p_1\) is the probability that thermal fluctuations drive the bond cluster to the stabilization threshold before they dissolve it completely. The mean absorption time is the characteristic time for this competition to resolve. The conditional time \(T_n^*\) is the mean time to stabilization given that the cluster succeeds. The Doob transform shows that conditioning on success is equivalent to a trajectory selection that suppresses complete bond loss; it is not a modification of the underlying binding free energy. This selection principle operates at the single-particle level. A particle that forms a first bond is not yet committed to internalization. The bond cluster must grow against a kinetic bias set by the ratio of dissociation to association rates. The probability \(p_1\) quantifies the chance that thermal fluctuations win this competition before all bonds are lost, and the conditional time \(T_n^*\) quantifies how long the successful trajectories take. These two quantities are not interchangeable: a virus or nanoparticle can have a high probability of eventual stabilization but require a long time to achieve it, or it can stabilize quickly only on rare successful attempts.

The rates \(\lambda_n\) and \(\mu_n\) are allowed to depend on \(n\), so the kinetic bias can change as the cluster grows. This is the mesoscopic signature of cooperative effects such as ligand mobility, configurational entropy, and steric interactions \cite{Kitov2003,ZhangLiWang2021PRE,ChenXueLiEtAl2024,DiMicheleJanaMognetti2018}. The birth--death chain does not resolve these effects microscopically, but it provides the exact probabilistic structure that any microscopic model satisfying the Markovian coarse-graining and time-scale separation must reduce to at the mesoscopic level. In this sense, the theory connects detailed molecular simulations of bond formation and continuum models of membrane wrapping.

The present theory addresses the discrete stage between the pre-contact Smoluchowski--Robin description of the companion paper \cite{Wu2026first} and the wrapping-stage Onsager description \cite{Liu2026e1,Liu2026e2}. The Smoluchowski--Robin theory ends at first-bond formation and does not resolve bond-number fluctuations. The Onsager theory assumes that contact is already established and compresses the binding into a continuous adhesion density \(a\). Neither theory provides the hitting probability, the mean absorption time, or the conditional stabilization time of the discrete bond cluster. The present theory supplies these quantities and states the conditions under which the discrete chain factorizes with the spatial capture probability. The factorization \(P_{\mathrm{stable}}=P_{\mathrm{enter}}u(d_0)p_1\) separates the probability of reaching the first bond from the probability that the bond cluster stabilizes. When the coupling between gap and bond rates is strong, the post-capture generator \eqref{eq:post-generator} must be embedded in a fully coupled model with the pre-capture transfer \eqref{eq:pre-transfer} or an appropriate generalization. This coupled generator is the discrete-bond analogue of the continuous spatial--mechanical coupling used in the Onsager wrapping theory.

A central question for comparison with measurement is which combinations of the rates can be inferred from observable quantities. The hitting probability \(p_1\) depends only on the ratios \(\rho_n=\mu_n/\lambda_n\), not on the absolute scale of the rates. More generally, the hitting probabilities are invariant under state-dependent positive rescalings
\[
\lambda_n\mapsto C_n\lambda_n,
\qquad
\mu_n\mapsto C_n\mu_n,
\qquad
C_n>0,
\]
because all local ratios \(\rho_n\) are unchanged. Under a common scaling \(\lambda_n'=c\lambda_n\), \(\mu_n'=c\mu_n\) with \(c>0\), all hitting probabilities are unchanged and all mean absorption times and conditional stabilization times are scaled by \(1/c\). For the constant-rate model, \(p_1=p_1(\rho,m)\). A single measured \(p_1\) gives one equation for two unknowns \((\rho,m)\) and does not identify either without additional structure. Even with \(m\) known, probability data identify \(\rho\) but not the common scale of \(\lambda\) and \(\mu\). To separate association and dissociation rates, one needs time-resolved bond-number trajectories, dwell times at intermediate states, or perturbation experiments that change \(\lambda\) and \(\mu\) differently. The Doob transform preserves the total exit rate,
\[
\lambda_n^*+\mu_n^*=\lambda_n+\mu_n,
\]
so the phrase that successful trajectories are faster should not be interpreted as an acceleration of the local waiting clock at each state. Conditioning reweights the jump directions and the path occupations, while the local exponential waiting time at a fixed state is unchanged. The model treats \(\lambda_n\) and \(\mu_n\) as mesoscopic parameters that absorb force dependence, rebinding, and cooperative interactions. The threshold \(m\) is a model definition rather than a directly measured ligand count. To interpret \(m\) as stable adhesion, one needs an independent argument that the escape time from \(m\) is long compared with the observation window or the subsequent membrane-wrapping time; otherwise the computed quantity is the probability of reaching the stabilization threshold, not a full experimental stable-adhesion probability. The factorization of the stable capture probability into \(u(d_0)p_1\) requires conditional independence and gap-independent bond kinetics. When spatial and bond degrees of freedom interact strongly, the post-capture generator \eqref{eq:post-generator} must be used inside a fully coupled model instead of the factorized product.

The present chain treats the membrane as a passive planar substrate and the bond rates as fixed mesoscopic parameters. Two extensions follow naturally. First, the membrane is deformable and multicomponent, so the local adhesion energy, receptor mobility, and bond kinetics depend on membrane composition, curvature, and tension. Coupling the bond-number chain to a Helfrich--Canham membrane \cite{helfrich1973,Canham1970,Ouyang1989,wu2023} with a multicomponent lipid \cite{wu2018getting,wu2025generalized} or receptor distribution would allow the stabilization threshold to depend on the local membrane state rather than on a fixed \(m\). Second, multivalent binding can produce superselective responses, in which the bound fraction changes sharply over a narrow range of receptor density or ligand affinity \cite{MartinezVeracoechea2011}. The present hitting probability is a monotone function of the kinetic bias ratio, but the superselective regime arises when the effective rates themselves depend on the number of available receptors and on the membrane configuration. 

Future work can extend the chain to state-dependent rates \(\lambda_n(d),\mu_n(d)\), to semi-Markov dwell times, or to a coupled spatial--bond boundary value problem with experimentally informed first-bond kinetics. Such extensions should preserve the probability--time distinction and the identifiability conditions emphasized here. The coupling to membrane deformation and cytoskeletal viscoelasticity, as treated in the Onsager wrapping theory \cite{Liu2026e1,Liu2026e2}, is a natural next step: the bond-number chain can provide the discrete adhesion statistics that enter the effective adhesion density of the continuum wrapping model. On the experimental side, time-resolved bond-number trajectories, dwell-time distributions, and perturbation experiments that change association and dissociation rates differently would provide the data needed to identify the absolute rate scale. Such data would allow quantitative tests of multivalent stabilization in viral entry, targeted drug delivery, and nanoparticle uptake.

\section{Conclusion}

A finite birth--death chain provides a minimal but exact kinetic description of the post-capture stage of multivalent adhesion. The hitting probability, the mean absorption time, and the conditional stabilization time answer three distinct physical questions: which boundary is reached first, how long the competition takes, and how long successful trajectories take. The Doob transform makes explicit the trajectory-selection effect of conditioning on success. The spatial--bond interface states when the stable capture probability factorizes into the product of the first-capture probability and the chain hitting probability, and when a fully coupled model must be used instead. The analysis shows that hitting probabilities constrain only the rate ratios, whereas the absolute rate scale requires time-resolved or dwell-time data. The threshold \(m\) is a model-defined stabilization criterion, not a measured ligand count. The rates and threshold are effective parameters specified independently or inferred from kinetic data. The theory describes the discrete stage between the pre-contact Smoluchowski--Robin description and the wrapping-stage Onsager description, and it provides a quantitative framework for interpreting attachment probabilities, lifetimes, and the transition from first bond formation to a model-defined stabilization threshold.

\begin{acknowledgments}
H.W. thanks J. Liu, X. Wang, and J. Xu for helpful discussions. This work is supported by the General Program of National Natural Science Foundation of China under Grant No. 12374210, the open research fund of Songshan Lake Materials Laboratory No. 2023SLABFN20, and the startup fund No. WIUCASQD2022005 from the Wenzhou Institute, University of Chinese Academy of Sciences.
\end{acknowledgments}


\begin{thebibliography}{99}

\bibitem{Doherty2009}
G.~J. Doherty and H.~T. McMahon,
Mechanisms of endocytosis,
Annu. Rev. Biochem. \textbf{78}, 857--902 (2009).

\bibitem{zhang2015physical}
S.~Zhang, H.~Gao, and G.~Bao,
Physical principles of nanoparticle cellular endocytosis,
ACS Nano \textbf{9}, 8655--8671 (2015).

\bibitem{wu2026unveiling}
H.~Wu and Z.-C. Ou-Yang,
Unveiling the heterogeneity and multifunctions of biological membranes: A unifying perspective in membrane biophysics from molecular coupling to cellular function,
Membranes \textbf{16}, 79 (2026).

\bibitem{Gao2005}
H.~Gao, W.~Shi, and L.~B. Freund,
Mechanics of receptor-mediated endocytosis,
Proc. Natl. Acad. Sci. USA \textbf{102}, 9469--9474 (2005).

\bibitem{Shi2008}
W.~Shi, J.~Wang, X.~Fan, and H.~Gao,
Size and shape effects on diffusion and absorption of colloidal particles near a partially absorbing sphere: Implications for uptake of nanoparticles in animal cells,
Phys. Rev. E \textbf{78}, 061914 (2008).

\bibitem{mercer2010virus}
J.~Mercer, M.~Schelhaas, and A.~Helenius,
Virus entry by endocytosis,
Annu. Rev. Biochem. \textbf{79}, 803--833 (2010).

\bibitem{Marsh2006}
M.~Marsh and A.~Helenius,
Virus entry: Open sesame,
Cell \textbf{124}, 729--740 (2006).

\bibitem{Wu2026first}
H.~Wu,
Before the First Bond: Depletion-Assisted Stochastic Capture at the Onset of Endocytosis,
arXiv preprint, arXiv:2611.01321 (2026).

\bibitem{Liu2026e1}
J. Liu, Z.-C. Ou-Yang, and H. Wu,
Depletion Capture and Rayleigh--Onsager Dynamics in Receptor-Mediated Endocytosis,
arXiv \textbf{2026}, arXiv:2607.12766.

\bibitem{Liu2026e2}
J. Liu, Z. Ouyang, and H. Wu,
Entropy-Driven Initiation and Cytoskeletal Viscoelasticity in Endocytosis: An Onsager Variational Framework,
Membranes \textbf{16}, 305 (2026).

\bibitem{Erdmann2004}
T.~Erdmann and U.~S. Schwarz,
Stochastic dynamics of adhesion clusters under shared constant force and with rebinding,
J. Chem. Phys. \textbf{121}, 8997--9017 (2004).

\bibitem{Asakura1954}
S.~Asakura and F.~Oosawa,
On interaction between two bodies immersed in a solution of macromolecules,
J. Chem. Phys. \textbf{22}, 1255--1256 (1954).

\bibitem{Asakura1958}
S.~Asakura and F.~Oosawa,
Interaction between particles suspended in solutions of macromolecules,
J. Polym. Sci. \textbf{33}, 183--192 (1958).

\bibitem{MendezAlcaraz2000}
J.~M. M\'endez-Alcaraz, M.~Ch\'avez-P\'aez, R.~Casta\~neda-Priego, and M.~Medina-Noyola,
Depletion forces in colloidal mixtures,
Phys. Rev. E \textbf{61}, 4095 (2000).

\bibitem{Vrij1976}
A.~Vrij,
Polymers at interfaces and the interactions in colloidal dispersions,
Pure Appl. Chem. \textbf{48}, 471--483 (1976).

\bibitem{Dinsmore1995}
A.~Dinsmore, A.~Yodh, and D.~Pine,
Phase behavior of a binary mixture of large and small spheres,
Phys. Rev. E \textbf{52}, 4045--4057 (1995).

\bibitem{Imhof1995}
A.~Imhof and J.~Dhont,
Experimental phase diagram of a binary colloidal hard-sphere mixture with a large size ratio,
Phys. Rev. Lett. \textbf{75}, 1662--1665 (1995).

\bibitem{Steiner1995}
U.~Steiner, A.~Meller, and J.~Stavans,
Entropy-driven phase separation in binary mixtures of colloids,
Phys. Rev. Lett. \textbf{74}, 4750--4753 (1995).

\bibitem{Ilett1995}
S.~M. Ilett, A.~Orrock, W.~Poon, and P.~Pusey,
Phase behavior of a model colloid--polymer mixture,
Phys. Rev. E \textbf{51}, 1344--1352 (1995).

\bibitem{Dinsmore1998}
A.~Dinsmore, D.~Wong, P.~Nelson, and A.~Yodh,
Hard spheres in vesicles: Curvature-induced forces and particle-induced curvature,
Phys. Rev. Lett. \textbf{80}, 409--412 (1998).

\bibitem{Irajizad2017}
E.~Irajizad, N.~Walani, S.~L. Veatch, A.~P. Liu, and A.~Agrawal,
Depletion forces drive membrane ruffling and protein clustering at the nano--bio interface,
Soft Matter \textbf{13}, 1455--1465 (2017).

\bibitem{Wan2024}
H.~Wan, D.~Xu, L.~Gao, and L.-T. Yan,
Entropy in the nano--bio interface: From depletion to crowding,
Small Sci. \textbf{4}, 2300078 (2024).

\bibitem{Batra2009}
J.~Batra, K.~Xu, S.~Qin, and H.~X. Zhou,
Effect of macromolecular crowding on protein binding stability: Modest stabilization and significant biological consequences,
Biophys. J. \textbf{97}, 906--911 (2009).

\bibitem{Gupta2017}
A.~N. Gupta and J.~R. van der Maarel,
Compaction of plasmid DNA by macromolecular crowding,
Macromolecules \textbf{50}, 1666--1671 (2017).

\bibitem{Dey2022}
D.~Dey, A.~Nunes-Alves, R.~C. Wade, and G.~Schreiber,
Diffusion of small molecule drugs is affected by surface interactions and crowder proteins,
iScience \textbf{25}, 105088 (2022).

\bibitem{ellis2001crowding}
R.~J. Ellis,
Macromolecular crowding: Obvious but underappreciated,
Trends Biochem. Sci. \textbf{26}, 597--604 (2001).

\bibitem{Bell1978}
G.~I. Bell,
Models for the specific adhesion of cells to cells,
Science \textbf{200}, 618--627 (1978).

\bibitem{EvansRitchie1997}
E.~Evans and K.~Ritchie,
Dynamic strength of molecular adhesion bonds,
Biophys. J. \textbf{72}, 1541--1555 (1997).

\bibitem{Kitov2003}
P.~I. Kitov and D.~R. Bundle,
On the nature of the multivalency effect: A thermodynamic model,
J. Am. Chem. Soc. \textbf{125}, 16271--16284 (2003).

\bibitem{ZhangLiWang2021PRE}
Y.~Zhang, L.~Li, and J.~Wang,
Tuning cellular uptake of nanoparticles via ligand density: Contribution of configurational entropy,
Phys. Rev. E \textbf{104}, 054405 (2021).

\bibitem{ChenXueLiEtAl2024}
Y.-Q. Chen, M.-D. Xue, J.-L. Li, D.~Huo, H.-M. Ding, and Y.~Ma,
Uncovering the importance of ligand mobility on cellular uptake of nanoparticles: Insights from experimental, computational, and theoretical investigations,
ACS Nano \textbf{18}, 6463--6476 (2024).

\bibitem{DiMicheleJanaMognetti2018}
L.~Di Michele, P.~K. Jana, and B.~M. Mognetti,
Steric interactions between mobile ligands facilitate complete wrapping in passive endocytosis,
Phys. Rev. E \textbf{98}, 032406 (2018).

\bibitem{Johnston2016}
S.~T. Johnston, R.~E. Baker, and M.~J. Simpson,
Filling the gaps: A robust description of adhesive birth-death-movement processes,
Phys. Rev. E \textbf{93}, 042413 (2016).

\bibitem{Billah2023}
M.~M. Billah, H.~Deng, P.~Dutta, and J.~Liu,
Effects of receptor properties on particle internalization through receptor-mediated endocytosis,
Soft Matter \textbf{19}, 5907--5915 (2023).

\bibitem{FedotovAlexandrov2024}
S.~Fedotov and D.~V. Alexandrov,
Model for random internalization of nanoparticles by cells,
Phys. Rev. E \textbf{110}, 044101 (2024).

\bibitem{Chetrite2015}
R.~Chetrite and H.~Touchette,
Variational and optimal control representations of conditioned and driven processes,
J. Stat. Mech. (2015) P12001.

\bibitem{Norris1997}
J.~R. Norris,
\emph{Markov Chains},
Cambridge University Press, Cambridge, 1997.

\bibitem{KarlinTaylor1975}
S.~Karlin and H.~M. Taylor,
\emph{A First Course in Stochastic Processes},
2nd ed., Academic Press, New York, 1975.

\bibitem{Gardiner2004}
C.~W. Gardiner,
\emph{Handbook of Stochastic Methods},
Springer, Berlin, 2004.

\bibitem{Risken1989}
H.~Risken,
\emph{The Fokker--Planck Equation: Methods of Solution and Applications},
2nd ed., Springer, Berlin, 1989.

\bibitem{Redner2001}
S.~Redner,
\emph{A Guide to First-Passage Processes},
Cambridge University Press, Cambridge, 2001.

\bibitem{helfrich1973}
W.~Helfrich,
Elastic properties of lipid bilayers: theory and possible experiments,
Z. Naturforsch. C \textbf{28}, 693--703 (1973).

\bibitem{Canham1970}
P.~B. Canham,
The minimum energy of bending as a possible explanation of the biconcave shape of the human red blood cell,
J. Theor. Biol. \textbf{26}, 61--81 (1970).

\bibitem{Ouyang1989}
Z.-C. Ou-Yang and W.~Helfrich,
Bending energy of vesicle membranes: general expressions for the first, second, and third variation of the shape energy and applications to spheres and cylinders,
Phys. Rev. A \textbf{39}, 5280--5288 (1989).

\bibitem{wu2023}
H.~Wu, Z.-C. Ou-Yang, and R.~Podgornik,
A note on vestigial osmotic pressure,
Membranes \textbf{13}, 332 (2023).

\bibitem{wu2018getting}
H.~Wu, M.~A. P. de Le\'on, and H.~G. Othmer,
Getting in shape and swimming: the role of cortical forces and membrane heterogeneity in eukaryotic cells,
J. Math. Biol. \textbf{77}, 595--626 (2018).

\bibitem{wu2025generalized}
H.~Wu and Z.-C. Ou-Yang,
A generalized Helfrich free energy framework for multicomponent fluid membranes,
Membranes \textbf{15}, 182 (2025).

\bibitem{MartinezVeracoechea2011}
F.~J. Martinez-Veracoechea and D.~Frenkel,
Designing super selectivity in multivalent nano-particle binding,
Proc. Natl. Acad. Sci. U.S.A. \textbf{108}, 10963--10967 (2011).

\end{thebibliography}
\end{document}